# Competition between spin-glass and antiferromagnetic states in Tsai-type 1/1 and 2/1 quasicrystal approximants

Farid Labib[1*], Hiroyuki Takakura[2], Asuka Ishikawa[3] and Ryuji Tamura[1]

[1] *Department of Material Science and Technology, Tokyo University of Science, Tokyo 125-8585, Japan*
[2] *Division of Applied Physics, Faculty of Engineering, Hokkaido University, Sapporo 060-8628, Japan*
[3] *Research Institute of Science and Technology, Tokyo University of Science, Tokyo 125-8585, Japan*

Systematic research was performed to investigate magnetic properties of the Tsai-type Ga-Pd-*RE* (*RE* = Gd, Tb, Dy, and Ho) systems, where both 1/1 and 2/1 quasicrystal approximants (ACs) are attainable at the same compositions as thermodynamical stable phases. Most of the samples exhibited spin-glass (SG)-like freezing behavior at low temperatures except Ga-Pd-Tb 2/1 AC and Ga-Pd-Ho 1/1 AC. The former showcased antiferromagnetic order at 5.78 K while the latter did not show any anomaly down to 1.8 K. Furthermore, 2/1 ACs were noticed to be less frustrated than their corresponding 1/1 ACs (based on the empirical rule of $|\theta_W/T_f|$) presumably due to the disorder-free environment in the nearest neighbour of the rare earth sites that form a network of distorted octahedra in the 2/1 ACs. The spin dynamic in SG samples was also characterized by means of ac magnetic susceptibility measurements. The results evidenced a weak response of the freezing temperatures to the measurement frequency in the Heisenberg systems, i.e., Gd-contained ACs, in contrast to the non-Heisenberg systems, i.e., Tb, Dy and Ho-contained ACs, where significant dependency is noticed for the latter. The spin-glass samples were further examined by fitting their freezing temperatures to the Vogel–Fulcher law.

## I. INTRODUCTION

Icosahedral quasicrystals (*i*QCs), as aperiodically-ordered intermetallic compounds, generate sharp Bragg reflections with 5-fold rotational symmetry indicating the presence of a long-range order without periodicity in their atomic configuration [1,2]. Their main structural building unit is a rhombic triacontahedron (RTH) cluster [3], which is a multi-shell polyhedron composed of four inner units (from the outermost one): an icosidodecahedron, an icosahedron, a dodecahedron, and a central unit, which is usually but not always a disordered tetrahedron [4]. The rare earth (*RE*) elements, as schematically shown in Fig. 1a, typically occupy the vertices of the icosahedron shell. Approximant crystals (ACs), on the other hand, refer to periodic counterparts of the *i*QCs. The term "AC" is preceded by a rational approximation $f_{n+1}/f_n$ of τ defined as $(1+\sqrt{5})/2$ where $f_n$ denotes the *n*-th Fibonacci number [2]. In this context, the higher the approximation we take, the more the structure resemblances to that of *i*QC [4]. Figures 1b and c represent the configuration of *RE* sites in the two lowest approximants, i.e., 1/1 and 2/1 ACs. Within one unit cell of 1/1 AC, as seen in Fig. 1b, there exist 24 symmetrically equivalent *RE* sites [4], which increases to 104 sites with 5 different symmetries in the 2/1 AC (represented by spheres of different colors in Fig. 1c) [5,6]. Among these sites, $RE_1-RE_4$ belong to icosahedron vertices and four $RE_5$ dimers locate on the body diagonal axes of Acute rhombohedron (AR) units that fill the vacant space in between the RTH clusters. More details about structure parameters of *i*QC and their ACs can be found elsewhere [3].

As far as their magnetism is concerned, most of the Tsai-type *i*QCs and ACs reported to date showcase canonical spin-glass (SG)-like freezing behavior [7,8] while some exhibit long-range ferromagnetic (FM) [9–14], and antiferromagnetic (AFM) [15–20] order. So far, all the experimental efforts in developing stable *i*QCs with a long-range magnetic order, especially of an AFM-type, have failed despite the claim of theoretical works, based on which no symmetry-related argument exists to prevent the establishment long-range magnetic order in *i*QCs [21–27]. This suggests that the SG state in *i*QCs is induced by rather non-symmetry-related parameters. One may consider structural defects such as chemical and/or positional disorders, random distribution of $RE - RE$ distances and the distribution of the easy axis or plane (in the non-Heisenberg members) as potential parameters that could provoke SG-like behavior in *i*QCs [28]. The appearance of SG behavior in the binary Cd-Gd *i*QC [29], where neither chemical disorder nor magnetic anisotropy exist, suggests the distribution of $RE - RE$ distances as a more dominant contributing factor to the spin freezing phenomenon in *i*QCs. Quite recently, though, a FM order has been discovered in rapidly-quenched Au−Ga−Gd and Au−Ga−Tb *i*QCs providing the first experimental evidence of a long-range magnetic order in *i*QCs [14]. As a significant accomplishment, a correlation between electron concentration of the Tsai-type compounds (mostly Au-based) and their magnetic ground state has been proposed [9,10,17,20,30], based on which magnetic orders can be attained at lower electron per atom (*e/a*) ratios than ∼ 1.81. Opposed to such *e/a* guideline, a new Tsai-type 2/1 AC with AFM order has been recently reported in the Ga-Pd-Tb system around the composition of $Ga_{50}Pd_{35.5}Tb_{14.5}$ [19] where *e/a* equals 1.93 marking the first ternary magnetically ordered Tsai-type compound with relatively large *e/a*. This was later followed by a comprehensive research performed by the present authors aiming to clarify phase stability and atomic structure of the Ga-Pd-*RE* (*RE* = Gd, Tb, Dy, and Ho) ACs [31,32]. The results revealed the possibility of obtaining 1/1 and 2/1 ACs with the same compositions by just altering the annealing temperature. Such a possibility provided a unique opportunity to study the effect of structural evolution towards quasiperiodicity on the physical properties of Ga-Pd-based ACs.

As such, the present study is set out to assess the effect of not only atomic structure but also the *RE* type on the magnetic properties of Ga-Pd-*RE* (*RE* = Gd, Tb, Dy, and Ho) 1/1 and 2/1 ACs. This paper is among a few published works comparing magnetic behaviors of lower and higher order ACs and/or *i*QCs, which is a prerequisite for understanding the nature of magnetism in quasiperiodic structures. For example, P. Wang *et al.* [33] compared magnetic properties of SG $Ag_{50}In_{36}Gd_{14}$ *i*QC and 1/1 AC using dc and ac magnetic susceptibility. They reported one-stage freezing of Gd spins at $T_f$ (freezing temperature) = 4.3 K in $Ag_{50}In_{36}Gd_{14}$ *i*QC but two-stage freezing phenomenon in the 1/1 AC with the same composition at $T_{f1}$ = 3.7 K and $T_{f2}$ = 2.4 K. In a different study [34], magnetic properties of Au-Al-Tm *i*QC and 1/1 AC were compared and SG-like behavior was reported for both compounds, although the geometrical frustration parameter was lower in the *i*QC sample. Comparative study of the magnetic properties of Cd-Mg-*RE* (*RE* = Gd, Tb, Dy, Ho, Er, and Tm) *i*QC, 2/1 and 1/1 ACs with SG behavior [6], on the other hand, revealed higher frustration in quasiperiodic structure than periodic ones assumably due to more random *RE*-*RE* distance distribution in *i*QCs.

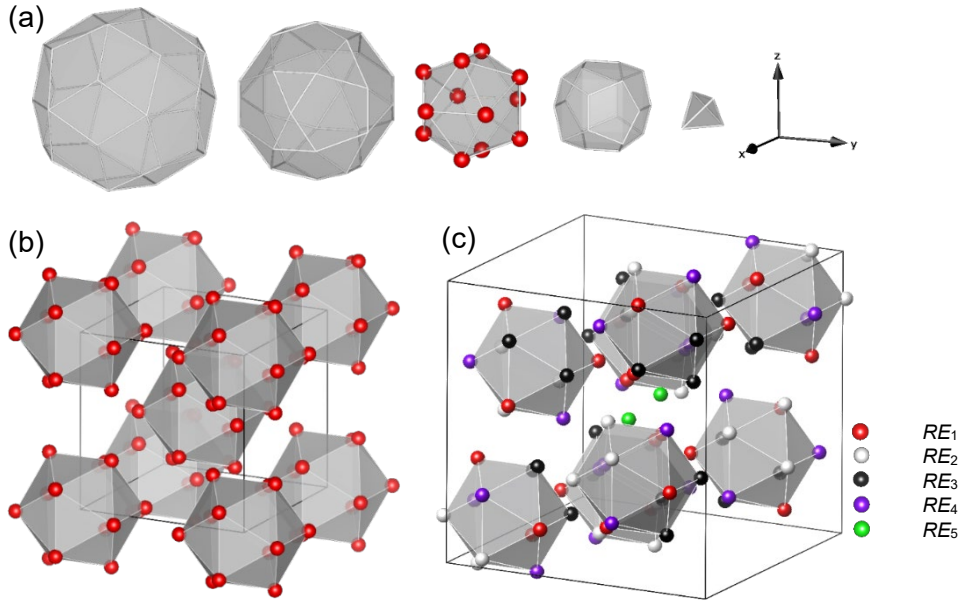

Fig. 1. (a) A typical shell structure of the main building unit in the Tsai-type icosahedral quasicrystals (*i*QCs). From left: a rhombic triacontahedron (RTH), an icosidodecahedron, an icosahedron, a dodecahedron and a tetrahedron. Depending on the alloy system, the inner tetrahedron could be either ordered or disordered. The *RE* occupies the vertices of the icosahedron. Configuration of *RE* sites within the unit cell of (b) 1/1 AC and (c) 2/1 AC. The spheres with different colors in Fig. 1c represent *RE* sites with five distinct symmetries, among which $RE_1$–$RE_4$ atoms occupy the icosahedron vertices and $RE_5$ atoms locate on the body diagonal axis of Acute Rhombohedron (AR) units that fill the vacant space in between the RTH clusters.

In this paper, we present experimental results of dc magnetic susceptibility of the Ga-Pd-*RE* (*RE* = Gd, Tb, Dy and Ho) 1/1 and 2/1 ACs. To confirm the SG state in some of the samples, we performed additional experiments such as ac magnetic susceptibility and metastable spin relaxation (or aging process). During the discussions, we seek to find possible connections between the atomic structure and the observed magnetic properties by referring to the refined structure models of the Ga-Pd-Tb 2/1 and 1/1 ACs provided elsewhere [31]. We will quantitatively analyze the frequency-dependence of the spin freezing temperature ($T_f$) in each of the studied SG samples based on the Vogel-Fulcher law [28,35], which will be explained in detail within the text. The findings of the paper should shed light on our understanding about the magnetism in aperiodic systems.

## II. EXPERIMENT

Polycrystalline alloys with nominal compositions of Gd, $Ga_{50}Pd_{36}RE_{14}$ (*RE* = Gd, Tb, Dy and Ho) were prepared from high purity elements using arc-melting and subsequent isothermal annealing at $T$ = 973 and 1073 K under Ar atmosphere for obtaining 2/1 and 1/1 ACs, respectively. Inductively Coupled Plasma (ICP) analysis of the samples, as listed in Table I, revealed a fair consistency with the nominal values. Powder X-ray diffraction (Rigaku SmartLab SE X-ray Diffractometer) with Cu-$K_\alpha$ radiation was performed for phase identification. The dc magnetic susceptibility of the samples was measured under zero-field-cooled (ZFC) and field-cooled (FC) modes using superconducting quantum interference device (SQUID) magnetometer (Quantum Design, MPMS3) in the temperature range from 1.8 K to 300 K and external dc fields up to $7 \times 10^4$ Oe. Further, ac magnetic susceptibility measurements were carried out under frequencies ranging from 0.1 to 100 Hz in the temperature range of 2 - 20 K and $H_{ac}$= 1 Oe. The time-dependence of the dc magnetic susceptibility (magnetic relaxation experiment) was measured in ZFC mode under 10 Oe for waiting times of $t_w$ = 100, 1000 and 5000 s.

## III. RESULTS

### III. 1. Phase identification

Figures 2a (resp. 2b) presents powder XRD patterns of the $Ga_{50}Pd_{36}RE_{14}$ (*RE* = Gd, Tb, Dy and Ho) compounds annealed at 973 K (resp. 1073 K) along with the results of Le Bail fittings [36] obtained by assuming the space groups $Pa\overline{3}$ (resp. $Im\overline{3}$) using the Jana 2006 software suite [37]. The red and black lines in the figure represent calculated ($I_{cal}$) and measured ($I_{obs}$) peak intensities, respectively, while the expected Bragg peak positions are shown by blue vertical bars. As shown, the experimental peak positions and their intensities are consistent with the calculation confirming the high purity of the synthesized 1/1 and 2/1 ACs.

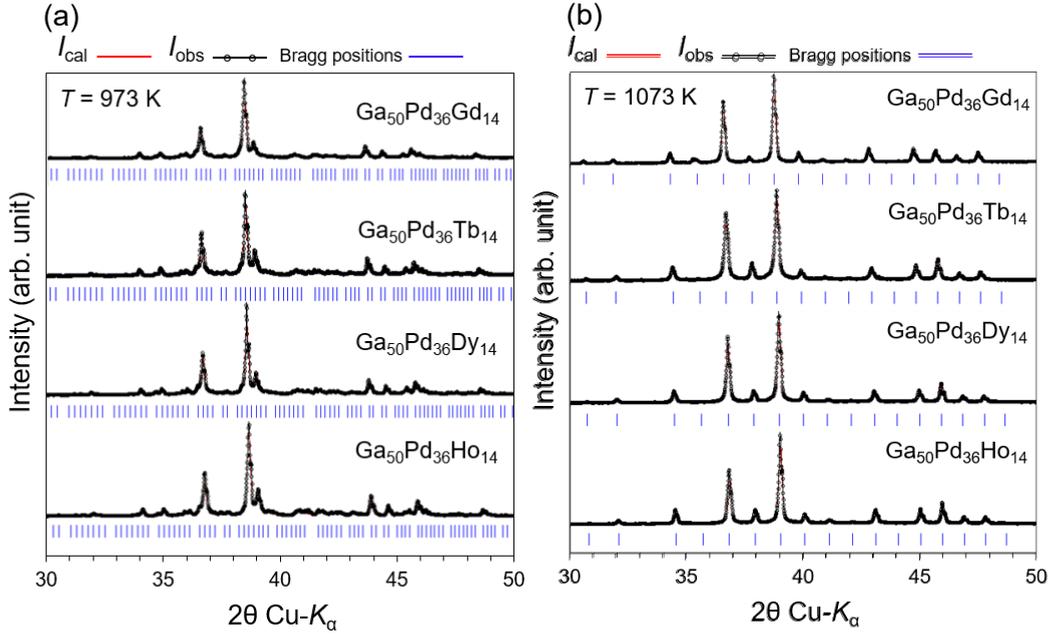

Fig. 2. The Le Bail fitting of powder x-ray diffraction (XRD) patterns of the $Ga_{50}Pd_{36}RE_{14}$ ($RE$ = Gd, Tb, Dy and Ho) compounds annealed at (a) 973 K and (b) 1073 K.

### III. 2. Magnetic properties (dc susceptibility)

The temperature-dependance of dc magnetic susceptibility of the $Ga_{50}Pd_{36}RE_{14}$ ($RE$ = Gd, Tb, Dy and Ho) 1/1 and 2/1 ACs at low temperatures ($T = 0 - 30$ K) in FC (filled circles) and ZFC (unfilled circles) modes are shown in the main panels of Fig. 3, where the 1/1 AC (resp. 2/1 AC) is represented by red (resp. blue). The high-temperature inverse magnetic susceptibility of all samples in the temperature range of 1.8–300 K (in the upper right insets of Fig. 3) exhibit a linear behavior which can be well fitted to the Curie-Weiss law defined as:

$$\chi(T) = \frac{N_A \mu_{eff}^2 \mu_B^2}{3k_B(T - \theta_w)} + \chi_0 \quad (1)$$

where $N_A$, $\mu_{eff}$, $\mu_B$, $k_B$, $\theta_w$ and $\chi_0$ denote Avogadro's number, effective magnetic moment, Bohr magneton, the Boltzmann factor, Curie-Weiss temperature (which indicates a sum of all exchange interactions) and the temperature-independent magnetic susceptibility, respectively [38]. The $\chi_0$ equals to almost zero in the present samples. The estimated $\theta_w$ (from linear fits to the inverse susceptibility in the temperature range of $T = 150 - 300$ K) and $\mu_{eff}$ are listed in Table I. All $\theta_w$ values are negative indicating a dominant AFM exchange interaction between the magnetic moments at high temperatures. The agreement between $\mu_{eff}$ and the theoretical magnetic moments of $RE^{3+}$ free ions defined as $g\sqrt{J(J+1)}\mu_B$ [38] suggests localization of the magnetic moments on $RE^{3+}$ ions.

At low temperatures, most samples except Ga-Pd-Tb 2/1 AC in Fig. 3b and Ga-Pd-Ho 1/1 AC in Fig.3d exhibit bifurcation between ZFC and FC curves below $T_f$ implying SG-like freezing behaviour. In the Ga-Pd-Tb 2/1 AC, both FC and ZFC curves display a sharp cusp at $T = 5.78$ K suggesting an AFM order establishment. The insets at the bottom-left corners of the main panels in Fig. 3 provide magnified views of the ZFC curves, where the positions of the cusps are more evident. Throughout the manuscript, the maximum temperature of ZFC magnetization is assumed as a $T_f$ (in the SG samples) and a Néel temperature $T_N$ (in the AFM Ga-Pd-Tb 2/1 AC). Notice that the Ho-contained 2/1 AC does not exhibit a maximum but a bifurcation between the FC and ZFC curves at $T$ = 3.2 K. Still, this temperature is assumed as a $T_f$ due to its nonequilibrium state evidenced by the appearance of frequency dependent cusp in the ac magnetic susceptibility at the same temperature (which will be discussed later in Fig. 8), as a hallmark of spin freezing phenomenon [39]. The estimated $T_f$ (or $T_N$) values are listed in Table I. The establishment of AFM order in the Ga-Pd-Tb 2/1 AC is further confirmed by observing metamagnetic-like anomalies in the field-dependance magnetization curves measured at $T$ = 1.8 K, 3 K and 5 K (all below the transition temperature at $T$ = 5.76 K), as displayed in Fig. 4. The first derivative curves in the inset associated with the maximum slope of the $M$-$H$ curves evidence the shift of the anomaly to lower fields (from $T$ = 0.91 T to 0.61 T) as the temperature raises from $T$ = 1.8 K to $T$ = 5 K. The occurrence of such anomaly and its shift to lower fields by increasing temperatures provides supporting evidence that the magnetic transition at $T$ = 5.76 K in the $Ga_{50}Pd_{36}Tb_{14}$ 2/1 AC is of AFM-type [15].

Table I. A list of analyzed composition (from ICP analysis), experimental Weiss temperature ($\Theta_w$), Néel temperature ($T_N$), freezing temperature ($T_f$), and effective magnetic moment ($\mu_{eff}$) obtained for Ga-Pd-$RE$ ($RE$ = Gd, Tb, Dy and Ho) 2/1 and 1/1 ACs

| AC type | Analyzed composition (ICP) | $\mu_{eff}$ ($\mu_B/RE_{ion}$) | $\mu_{calc.}$ ($\mu_B/RE_{ion}$) | $\Theta_w$ (K) | $T_{f1}$ (K) | $T_{f2}$ (K) | $T_N$ (K) | $|\Theta_w / T_f|$ |
|---|---|---|---|---|---|---|---|---|
| 1/1 AC | $Ga_{50.0}Pd_{35.8}Gd_{14.2}$ | 8.0±0.2 | 7.94 | -12.72±0.64 | 2.73 | – | – | 4.65 |
| 2/1 AC | $Ga_{50.1}Pd_{35.8}Gd_{14.1}$ | 8.2±0.2 | | -13.96±0.93 | 4.67 | – | – | 2.99 |
| 1/1 AC | $Ga_{49.0}Pd_{37.3}Tb_{13.7}$ | 9.8±0.1 | 9.72 | -10.32±0.89 | 3.22 | – | – | 3.20 |
| 2/1 AC | $Ga_{49.5}Pd_{36.8}Tb_{13.7}$ | 9.9±0.2 | | -10.13±1.26 | 5.78 | – | 3.4 | 1.75 |
| 1/1 AC | $Ga_{48.2}Pd_{37.2}Dy_{14.6}$ | 10.9±0.2 | 10.63 | -6.47±1.36 | 2.21 | – | – | 2.92 |
| 2/1 AC | $Ga_{48.6}Pd_{36.9}Dy_{14.6}$ | 10.9±0.3 | | -4.78±1.18 | 4.18 | 2.21 | – | 1.14 |
| 1/1 AC | $Ga_{48.5}Pd_{37.2}Ho_{14.3}$ | 10.8±0.2 | 10.58 | -2.25±0.84 | – | – | – | – |
| 2/1 AC | $Ga_{48.3}Pd_{37.2}Ho_{14.5}$ | 10.6±0.1 | | -1.53±0.97 | 3.2 | – | – | – |

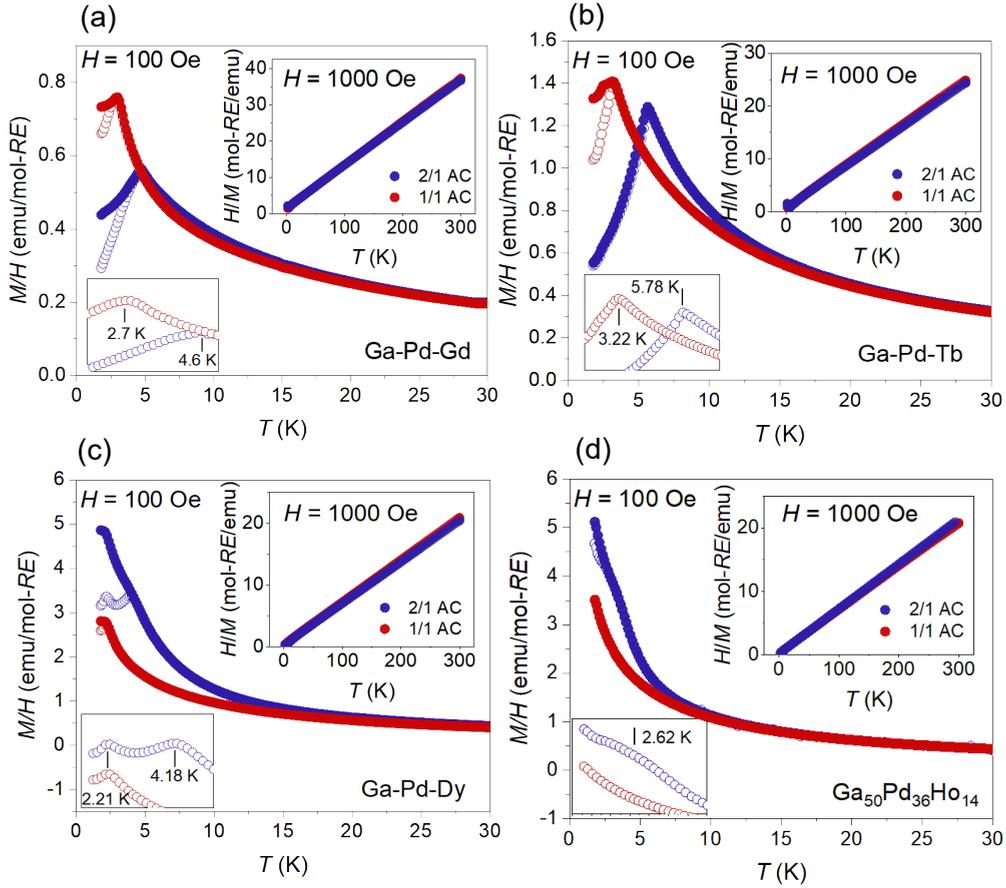

Fig. 3. Temperature-dependence of dc magnetic susceptibility of the $Ga_{50}Pd_{36}RE_{14}$ ($RE$ = Gd, Tb, Dy and Ho) 1/1 and 2/1 ACs under field-cooled (FC) and zero-field-cooled (ZFC) modes. The insets show corresponding inverse magnetic susceptibility results.

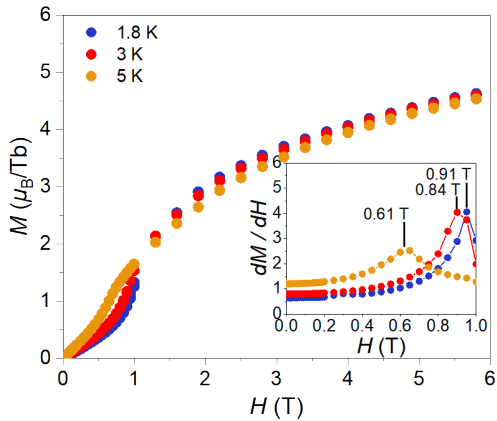

Fig. 4. Magnetization curves of the $Ga_{50}Pd_{36}Tb_{14}$ 2/1 AC as a function of applied magnetic field. The occurrence of metamagnetic-like anomalies is noticed through the maximum slope of the $M$-$H$ curve evidenced by sudden peaks in the first derivative curves in the inset. The position of peaks shifts to lower fields (from 0.91 T to 0.61 T) as the temperature raises from $T$ = 1.8 K to $T$ = 5 K.

Figures 5a and b depict the estimated $|\theta_w|$ and $T_f$ (or $T_N$) versus de Gennes parameter defined as $dG = (g - 1)^2 J(J + 1)$, where $g$ and $J$ denote the Landé g-factor and the total angular momentum, respectively. The 1/1 and 2/1 ACs are represented by red and blue filled circles, respectively (only AFM Ga-Pd-Tb 2/1 AC is shown by an unfilled circle for distinction). Figure 5a evidences approximate proportionality of $|\theta_w|$ to $dG$ indicating the domination of interaction in these compounds [40]. In Fig. 5b, however, a serious deviation of $T_f$ in Gd-contained ACs from $dG$ scaling, is observed, which is consistent with the previous studies [6,40–43] and is likely to originate from an isotropic behaviour of $Gd^{3+}$ ion in the crystalline electric field (CEF). It is a widely held view that the spin freezing

…

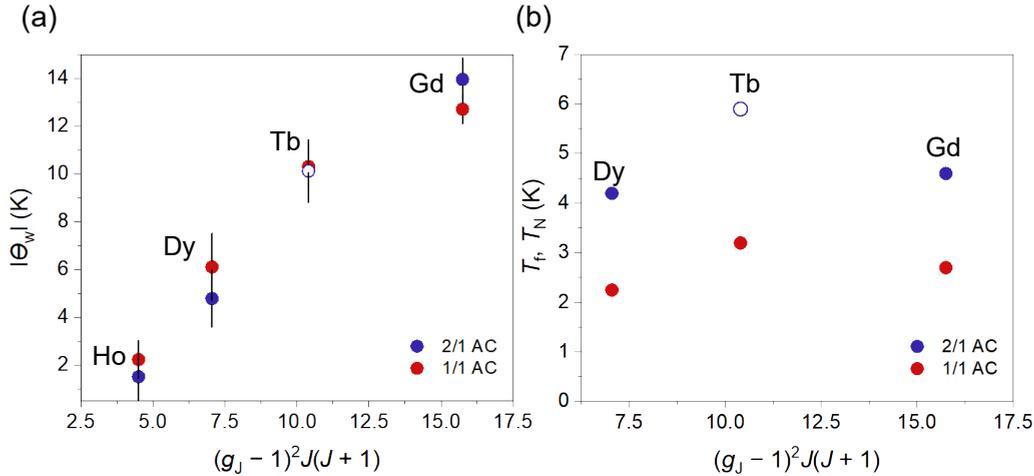

Fig. 5. Variation of (a) $|\theta_w|$ and (b) $T_f$ or $T_N$ as a function of the de Gennes factor: $dG = (g_J - 1)^2 J(J + 1)$ for the magnetic 2/1 and 1/1 ACs in the Ga-Pd-$RE$ systems. The values related to the AFM Ga-Pd-Tb 2/1 AC are represented by open circles in the figures. The relative correspondence between the experimental $|\theta_w|$ values and the $dG$ of the magnetic $RE$ is noticed in all samples except in the Gd-contained ones whose $|\theta_w|$ values are lower than expected possibly due to the isotropic behavior of $Gd^{3+}$ in the crystalline electric field.

phenomenon in the non-Heisenberg systems is governed by distribution of $RE$-$RE$ distances and easy axis (or plane) directions [28,40]. Since the latter is absent in the Heisenberg systems, their spin freezing phenomenon is considered to solely arise from the former leading to lower $T_f$ values than expected from $dG$ scaling. Comprehensive discussion in this subject is provided elsewhere [40,43].

Based on an empirical frustration parameter of $|\theta_w/T_f|$ [33,34,44], the frustration level is about two times lower in the 2/1 ACs than their corresponding 1/1 ACs (see Table I for the $|\theta_w/T_f|$ values). In principle, geometrical frustration arises when the lowest energy configuration of all two-spin interactions cannot be achieved at the same time due to competition with the neighboring spins. In particular, Tsai-type compounds wherein the $RE$ sites form a network of corner-sharing octahedra with the number of nearest-neighbor: $z = 8$ are considered as new type of frustrated magnets. The $|\theta_w/T_c|$ values in Table I indicate a weaker competition between spins in the 2/1 ACs with the space group $Pa\overline{3}$ compared to their corresponding 1/1 ACs with the space group $Im\overline{3}$. Figure 6a plots networks of Tb octahedra in the $Ga_{50}Pd_{35.5}Tb_{14.5}$ 1/1 and 2/1 ACs based on the diffraction data obtained from single-crystal X-ray measurement [31]. In Fig. 6b, the histogram distribution of the edge lengths of an isolated octahedron in the two ACs are compared. As shown, while the octahedron in the 1/1 AC is fairly symmetric evidenced by two distinct distances at 5.31 Å and 7.52 Å corresponding to its edges and body diagonals, respectively, the octahedron in the 2/1 AC is quite distorted evidenced by the widened distances distributed over a range of 0.35–0.55 Å. The appearance of malformed octahedra in the 2/1 AC, which was expected due to the existence of five different symmetrically equivalent $RE$ sites in it (see Fig. 1c), may partially relieve geometrical frustration inherent to the perfect octahedron spin arrangements provided that the spins are of Ising type.

Another possible contributor to different frustration levels in the 2/1 and 1/1 ACs could be the existence of structural defects such as fractionally occupied sites in the nearest neighbors of $RE$ positions in the structure of 1/1 AC that could locally break the mirror

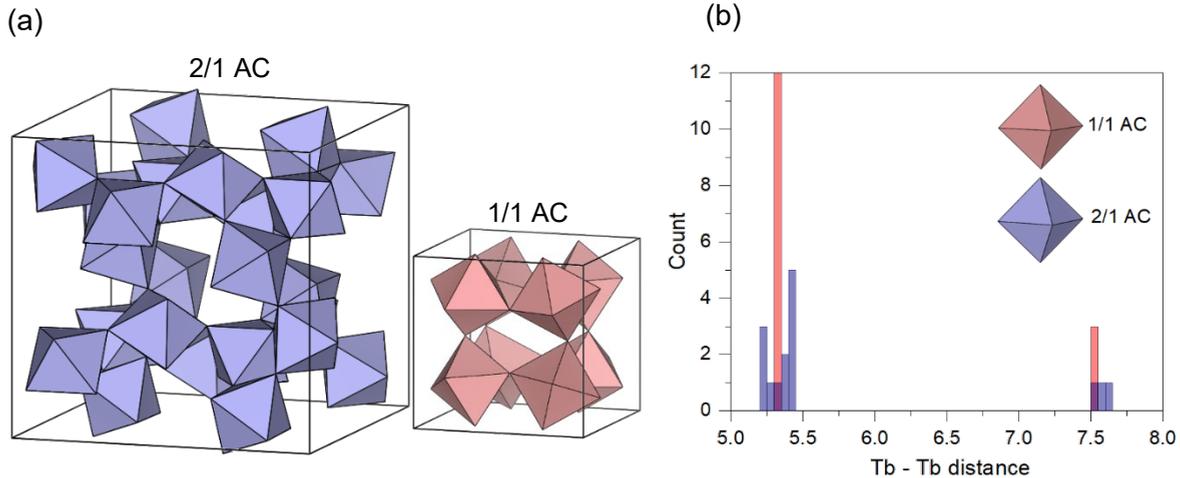

Fig. 6. (a) A network of corner-sharing tilted octahedra in the $Ga_{50}Pd_{35.5}Tb_{14.5}$ 1/1 and 2/1 ACs, (b) The histogram distribution of the edge lengths of an isolated octahedron in the two ACs.

…

symmetry and lead to different CEF potential for different *RE* sites and finally induce frustration. Figure 7 displays refined atomic structures within the nearest neighbor of an isolated $Tb^{3+}$ ion located at (a,b) icosahedron vertices and (c) along the long body diagonal of the AR in the $Ga_{50}Pd_{35.5}Tb_{14.5}$ (a) 1/1 AC and (b,c) 2/1 AC. Here, Ga, Pd and Tb atoms are represented by dark blue, light blue and red colours, respectively. What stands out in the figure is a significant positional and chemical disorder between Ga and Pd species that occurs on five out of sixteen vertices of the mono-capped, double, pentagonal antiprism polyhedron (see Fig. 7a) around $Tb^{3+}$ ions in the 1/1 AC. It can be argued that the localized 4*f* electrons of the $Tb^{3+}$ ions at the center of polyhedron are subject to the point charges of the Ga/Pd on the disordered sites through CEF effect which further induces spin frustration in the 1/1 AC. However, in the 2/1 AC, the vertices of $(Ga:Pd)_{18}Tb$ polyhedron around $Tb^{3+}$ ion on the icosahedron vertices (see Fig. 7b) and the AR unit that cages a Tb dimer outside the icosahedron shell (see Fig. 7c) are fully occupied by either Ga or Pd atoms forming highly ordered environments, which could be one of the contributing parameters in the establishment of AFM order in the Ga-Pd-Tb 2/1 AC. Note that the estimated frustration parameters in the present compounds are in the range of 1.14 – 4.65 being comparable to 2.3 – 3.6 in the ternary Au-Al-Tm [34], and 4.5 in ternary Zn-Mg-Tb [43,45], but lower than 8.7 and 15.5 in the Ag-In-Gd *i*QC and 1/1 AC [33], respectively, and 10 in binary *i*-Cd-Gd [8] systems. By definition, systems with frustration parameter of higher than 10 are classified as strongly frustrated systems [46], therefore, the present compounds can be considered as weakly frustrated compounds.

## III. 3. Magnetic properties (ac susceptibility)

One of the standard experiments to determine 'canonical' SG is through the ac magnetic susceptibility measurement [47]. Due to the non-equilibrium nature of SG systems [28], the magnitude and position of their $T_f$ should depend on the measurement frequency (*f*). Basically, the ac magnetic susceptibility has two components that are related through a relaxation time: in-phase or dispersion ($\chi'_{ac}$) and out-of-phase or absorption ($\chi''_{ac}$) components [28]. In the SG systems, the absorption effect due to the decoupling of spins from the lattice via relaxation process appears as a jump near $T_f$ in the $\chi''_{ac}$. The main panels in Fig. 8 depict temperature-dependence of $\chi'_{ac}$ for all samples (except Ga-Pd-Ho 1/1 wherein no anomaly is noticed down to 1.8 K) under *f* spanning three orders of magnitude from 0.1 to 100 Hz. The corresponding $\chi''_{ac}$ are provided in the inset. As shown, $\chi'_{ac}$ in some samples is frequency dependent below $T_f$ even though the level of dependency varies significantly with the *RE* species. At the same time, their corresponding $\chi''_{ac}$ rise from a zero background (again by different extents depending on the *RE* type) below $T_f$. The results, as summarized in table II, indicate 2 - 25 % shift of $T_f$ in the ACs by three decades variation of *f*. Figure 9 provides more clear perspective of $T_f$ variation in different *RE*-contained ACs by depicting the normalized $T_f$ (i.e., $T_f/T_{f\,(0.1\ Hz)}$) against *f* (in logarithmic scale). Take Ga-Pd-Ho 1/1 AC as an example, wherein the position of $T_f$ varies ∼ 15% by three orders of magnitude rise in *f*. The highest dependency belongs to $Ga_{50}Pd_{36}Dy_{14}$ 2/1 AC (Fig. 8f) wherein ∼25 % shift in the position of $T_f$ is noticed.

On the contrary, Gd-contained ACs (Figs. 8a and b) show the least sensitivity to the change of *f* as their $T_f$ varies merely by 1–2 % despite a clear divergence between their FC and ZFC magnetizations shown in Fig. 3a. Given that a similar trend in response of the Heisenberg and non-Heisenberg SGs to the frequency change has been formerly reported in other alloy systems such as Zn-Mg-*RE* (*RE* = Gd, Tb and Dy) *i*QCs [48] and even in non-QC-related SGs such as $B_{66}RE$ (*RE* = Gd, Tb, Ho, Er) [49], it seems unlikely to arise from an inherent nature of the present system. It is suggested to arise from magnetic anisotropy (either in easy axis or plane) energy, i. e., energy of the spin-orbital coupling via CEF, which is relatively weak in $Gd^{3+}$ with half-filled 4*f* shell (∼ 0.015 eV per atom [50]) and thus easier to be overcome during the spin rotation. Strong dependency of $\chi'_{ac}$ and $\chi''_{ac}$ susceptibilities on magnetic anisotropy and domain wall energies is a well-acknowledged phenomenon dealt by a number of theoretical and empirical studies [55–58].

As further experimental evidence for a glassy behavior in SG samples, ZFC dc magnetization of the samples under probing field of $H = 10$ Oe and $T = 2$ K is recorded after different waiting times of $t_w = 100$, 1000 and 5000 s (i.e., aging experiment). Figure 10 shows the magnetization versus $t_w$ (in logarithmic scale) for all samples. Overall, the dependency of the magnetization on the elapsed time $t_w$

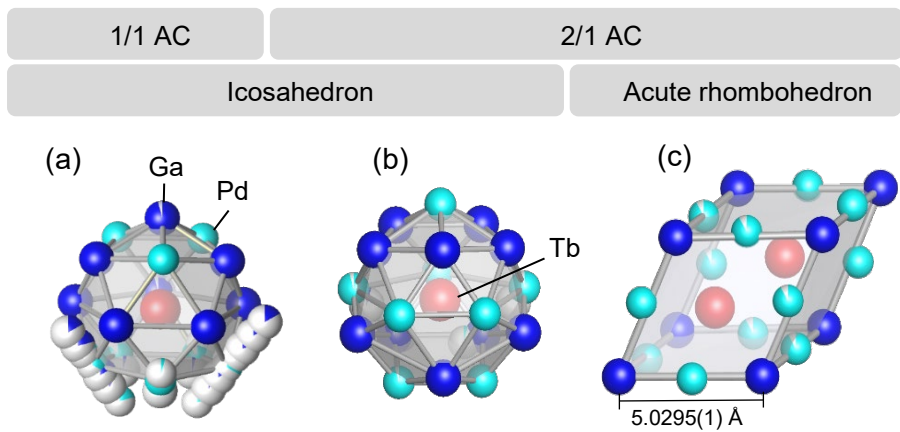

| 1/1 AC | 2/1 AC |
|---|---|
| Icosahedron | Acute rhombohedron |

Fig. 7. (a-c) Refined atomic structures within the first nearest neighbor of an isolated $Tb^{3+}$ ion on the (a,b) icosahedron vertices and (c) along the long body diagonal of acute rhombohedron in the Ga-Pd-Tb (a) 1/1 AC and (b,c) 2/1 AC. The Ga, Pd and Tb atoms are represented by dark blue, light blue and red colours, respectively.

…

Table II. A list of experimentally derived $T_f$, relative change in $T_f$ per decade change in $f$ K, Vogel–Fulcher temperature ($T_0$), $T_f - T_0$, and activation energy ($E_a/k_B$) for Ga-Pd-$RE$ ($RE$ = Gd, Tb, Dy and Ho) 2/1 and 1/1 ACs

| $RE$ type | 1/1 / 2/1 AC | $T_f(0.1)$ (K) | $T_f(1)$ (K) | $T_f(10)$ (K) | $T_f(100)$ (K) | $K$ | $T_0$ (K) | $T_f - T_0$ (K) | $E_a / k_B$ |
|---|---|---|---|---|---|---|---|---|---|
| Gd | 1/1 AC | 2.93 | 2.95 | 2.97 | 2.99 | 0.006(7) | 2.76(5) | 0.19(5) | 6.47±0.30 |
|  | 2/1 AC | 4.67 | 4.69 | 4.71 | 4.73 | 0.004(2) | 4.43(0) | 0.27(0) | 5.24±0.18 |
| Tb | 1/1 AC | 3.28 | 3.40 | 3.55 | 3.71 | 0.040(2) | 1.73(4) | 1.76(5) | 49.97±1.22 |
|  | 2/1 AC | – | – | – | – | – | – | – | – |
| Dy | 1/1 AC | 2.23 | 2.36 | 2.53 | 2.75 | 0.049(8) | 0.30(6) | 2.19(0) | 61.62±1.92 |
|  | 2/1 AC | 4.29 | 4.47 | 4.71 | 4.92 | 0.045(7) | 1.96(8) | 2.64(5) | 75.19±2.31 |
|  |  | 2.21 | 2.36 | 2.53 | 2.75 | 0.074(2) | 0.19(2) | 2.29(0) | 64.63±0.83 |
| Ho | 1/1 AC | – | – | – | – | – | – | – | – |
|  | 2/1 AC | 3.22 | 3.31 | 3.42 | 3.52 | 0.028(7) | 2.11(5) | 1.26(3) | 35.72±1.51 |

before the application of the magnetic field is noticed in SG samples even though it is more evident in the 1/1 ACs than their 2/1 counterparts assumably due to a pronounced chemical disorder and/or geometrical spin frustration inherent to the perfect $RE$ octahedra decoration in the structure of 1/1 ACs, as discussed earlier. Such phenomenon is a direct manifestation of the non-equilibrium nature of SG systems, the properties of which should depend on previous history of the system such as waiting time $t_w$ in a way that longer $t_w$ result in slower relaxation as the system gets "aged" during the waiting before application of the magnetic field. The Ga-Pd-Tb 2/1 AC, as expected, does not exhibit aging properties as it is an equilibrium phase. The Ga-Pd-Ho 2/1 AC, to our surprise, does not age after even $t_w = 5000s$. Looking at the $\chi'_{ac}$ and $\chi''_{ac}$ magnetizations in Fig. 8g, where convergence of magnetization curves under different frequencies at temperatures close to 1.8 K can be noticed, we suspect that an equilibrium state at some form is established below $T = 1.8$ K, which is the lowest temperature attainable experimentally using the available MPMS system. The clarification of this issue is left for future studies.

For a quantitative analysis of the results obtained from the ac magnetic susceptibility measurements, the first step is to calculate a relative change of $T_f$ per decade change in $f$ expressed as $K = \Delta T_f / (T_f \log f)$ for the SG samples, wherein $T_f$ indicates its average value over different frequencies. For Gd-contained 1/1 and 2/1 ACs, the obtained K values, as listed in Table Ⅱ, are in the range of $0.004(2) < K < 0.006(7)$ which is consistent with typical canonical SG systems such as $Cu_{1-x}Mn_x$ ($K = 0.005$) [28] and a factor of two lower than that in the $Ag_{50}In_{36}Gd_{14}$ $iQC$ ($K = 0.010$) [33]. For Tb-, Dy- and Ho-contained SG systems, the obtained K are in the range of $0.028(7)$ $K < 0.074(2)$, approximately one order of magnitude larger than those in the $Ga_{50}Pd_{36}Gd_{14}$ ACs, about three to seven times larger than those in the $Al_{1-x}Fe_x$ [28] and $Pd_{1-x}Mn_x$ ($K = 0.013$) [33] SGs and quite comparable with the $Zn_{57}Mg_{34}Tb_9$ $iQC$ ($K = 0.049$) [43]. Basically, the $K$ is a quantitative parameter that reflects the dependency level of $T_f$ to the measurement frequency.

Among the studied samples, the $Ga_{50}Pd_{36}Dy_{14}$ 2/1 AC exceptionally exhibits two distinct anomalies at $T_{f1} = 2.21$ K and $T_{f2} = 4.18$ K with both $\chi'_{ac}$ and $\chi''_{ac}$ being frequency dependent below $T_f$ (see Figs. 8e and f). Their shift by three orders of magnitude frequency change reaches up to nearly 25% and 15%, respectively. These features together with the occurrence of frequency-dependant peaks in the $\chi'_{ac}$ around $T_{f1}$ and $T_{f2}$ indicate two-stage relaxation process with different characteristic relaxation times which has also been formerly observed in a number of SG systems such as Ag-In-Gd 1/1 AC [33] and Cd-Mg-Tb 1/1 AC [59] as well as other geometrically frustrated magnets like $Gd_3Ga_5O_{12}$ [60], $Dy_{2-x}Yb_xTi_2O_7$ [61], $Dy_2Ti_2O_7$ [62] and $Fe_{1/4}TiS_2$ [63] and been correlated to formation of different magnetic clusters during the cooling process [63]. Almost a factor of two larger $K_1 = 0.071$ compared to $K_2 = 0.044$ in the $Ga_{50}Pd_{36}Dy_{14}$ 2/1 AC indicates that the spin relaxation at $T_{f1}$ is about two times faster than that at $T_{f2}$. As a direct confirmation of such implication, we measured the time-dependence spin relaxation (also called an aging process [28]) in the $Ga_{50}Pd_{36}Dy_{14}$ 2/1 AC at 1.8 K (below $T_{f1} = 2.21$ K), at 3.2 K (above $T_{f1} = 2.21$ K and below $T_{f2} = 4.18$ K), at 5.2 K and 10 K (both above $T_{f1}$ and $T_{f2}$) under 10 Oe. The results are shown in Fig. 11 wherein the normalized dc ZFC magnetization, i.e., $M(t) / M(t = 0)$, is depicted. Clearly, the magnetization at 1.8 K is less dependent on time than that at 3.2 K and is almost independent from time at 5.2 K and 10 K confirming different spin dynamics at each freezing stage in the $Ga_{50}Pd_{36}Dy_{14}$ 2/1 AC.

A final step for safe conclusion of SG state is to rule out the possibility of superparamagnet- type blocking in the present samples. One of the quantitative approaches in this direction is to fit the frequency dependence $T_f$ data to the following Arrhenius law:

$$f = f_0 \exp\left[-\frac{E_a}{k_B(T)}\right] \quad (2)$$

where $f$ denotes the driving force of the ac measurement and $T$, for our purpose, is considered as $T_f$. The best fitting of the data to Eqn. 2 yields completely unphysical values such as $f_0 = 10^{142}$ Hz and $E_a/k_B$ = 1009.4 K, as in the case of Ga-Pd-Tb 1/1 AC. The emergence of such unphysical fitting values, indeed, distinguishes the present SGs from superparamagnets where the Arrhenius law holds true [47]. The alternative approach for fitting the frequency-dependence of $T_f$ data is based on an empirical Vogel-Fulcher law which describes the viscosity of supercooled liquids in real glasses [28,35,64]. The Vogel-Fulcher law is described as below:

$$f = f_0 \exp\left[-\frac{E_a}{k_B(T_f - T_0)}\right] \quad (3)$$

…

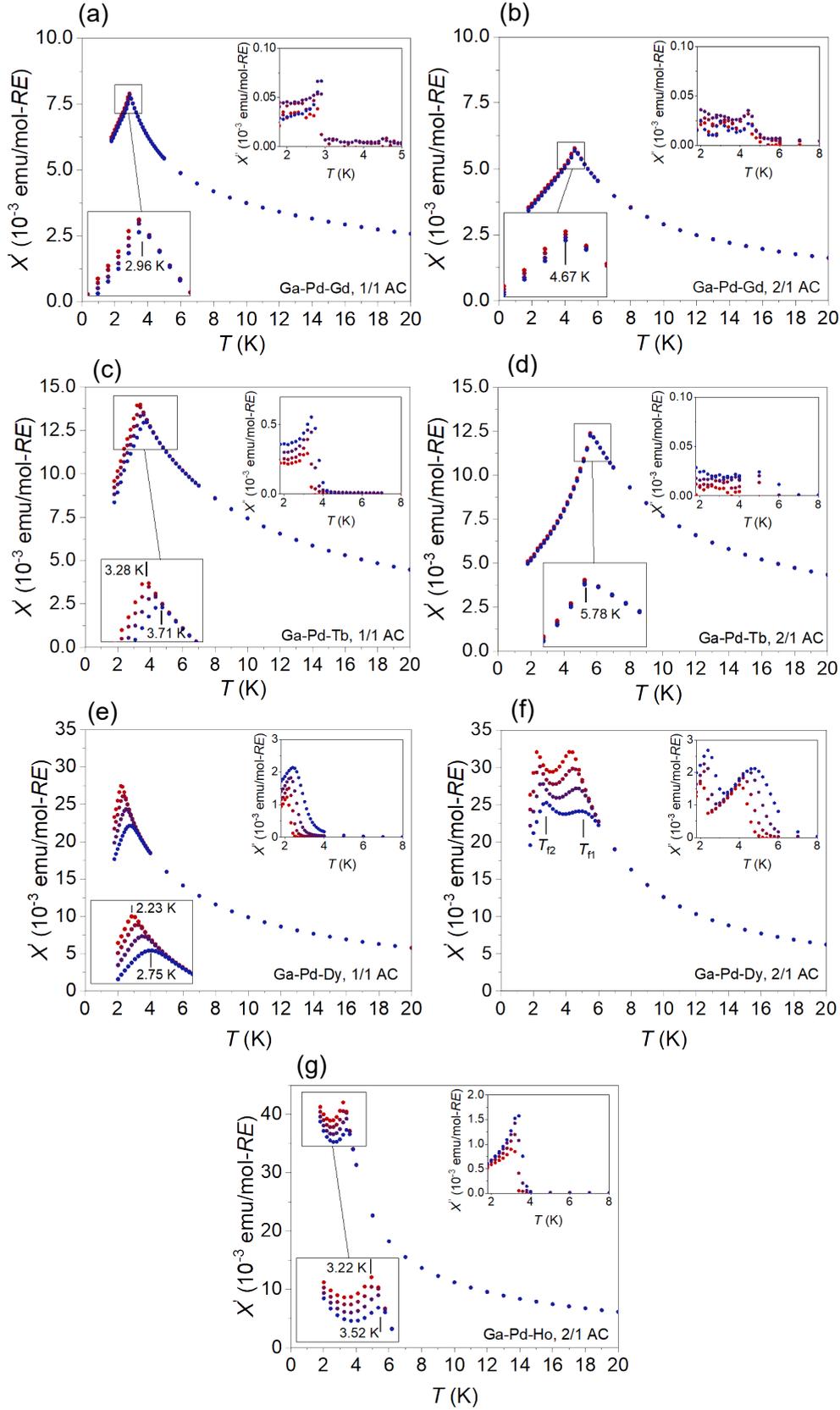

Fig. 8. The in-phase component of the ac susceptibility obtained from (a) Ga-Pd-Gd 1/1 AC, (b) Ga-Pd-Gd 2/1 AC, (c) Ga-Pd-Tb 1/1 AC, (d) Ga-Pd-Tb 2/1 AC, (e) Ga-Pd-Dy 1/1 AC, (f) Ga-Pd-Dy 2/1, and (g) Ga-Pd-Ho 2/1 within $1.8\ K < T < 20\ K$ under $f_{ac} = 0.1$ - $100$ Hz. The insets represent the corresponding out-of-phase components within $1.8\ K < T < 10\ K$.

...

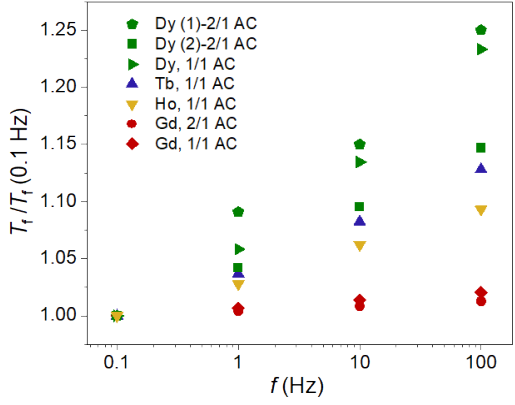

Fig. 9. Normalized $T_f$ by their lowest values obtained at 0.1 Hz versus the measurement $f$ for the present 1/1 and 2/1 ACs.

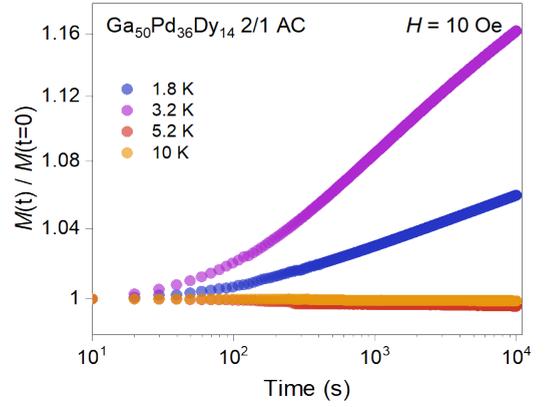

Fig. 11. Variation of the normalized dc magnetization $M(t)/M(t = 0)$ in ZFC mode versus time in logarithmic scale measured for the $Ga_{50}Pd_{36}Dy_{14}$ 2/1 AC at 1.8, 3.2 ,5.2 and 10 K under 10 Oe.

where, the $f_0$, $E_a$ and $T_0$ are fitting parameters. The latter is referred as an 'ideal glass temperature' in real glasses but in SGs, it is commonly considered as a measure of inter-cluster interaction strength [28] given that SG is a collection of interacting magnetic clusters after all. Some [35], however, correlate $T_0$ to a real temperature where an underlying phase transition, for which the $T_f$ obtained from dc measurement is just a dynamic manifestation, takes place. By assuming $f_0 = 1 \times 10^{13}$ Hz, as a typical value in SG systems [35], the best fit of the $T_f$ ($f$) to the Eqn. 3 results in $E_a/k_B$ and $T_0$ values listed in Table II. As seen, the $T_0$ in each compound is always lower than the corresponding $T_f$, which is a logical outcome if $T_0$ is considered as a real transition temperature. Figure 12 depicts variation of $1/(T_f - T_0)$ versus $f$ (in logarithmic scale) wherein a linear correspondence between the two is clear. Using the Vogel-Fulcher law, therefore, allows a reasonable interpretation of the frequency-dependence of $T_f$ with physical fitting parameters. Based on a number of reports [33,35,43,64–66], the fitting parameters $E_a/k_B$ and $T_0$ can further be used to derive more information about the nature SG. It has been proposed that in SGs with strong RKKY-type interaction, the $T_f - T_0 \ll T_f$ condition is satisfied, while for those with a weak dominance of RKKY interaction, one could expect $T_0 \ll T_f$. In the present ACs, the $T_f - T_0$ values are in a range of 0.19(5) – 0.27(0) in the Gd-contained ACs, 1.76(5) in the Tb-contained 1/1 AC,

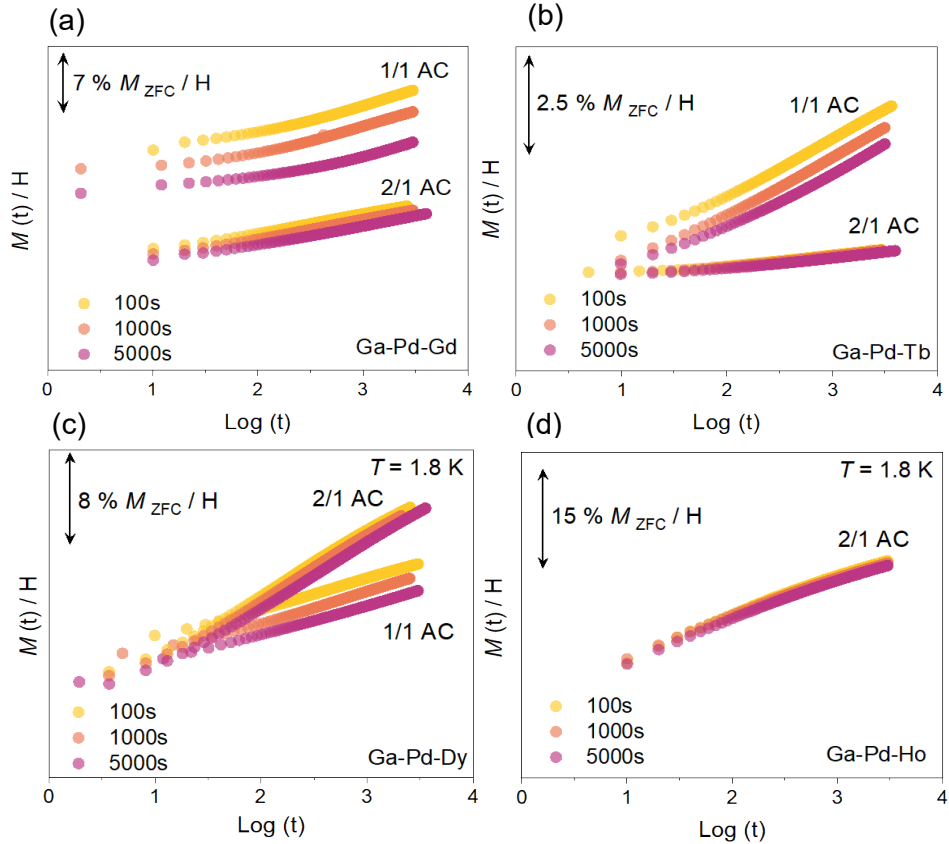

Fig. 10. ZFC magnetization of the $Ga_{50}Pd_{36}RE_{14}$ ($RE$ = Gd, Tb, Dy and Ho) 1/1 and 2/1 ACs ($RE$ = (a) Gd, (b) Tb, (c) Dy and (d) Ho) after isothermal aging at 1.8 K for different wait times of $t_w$ = 100, 1000 and 5000 s.

…

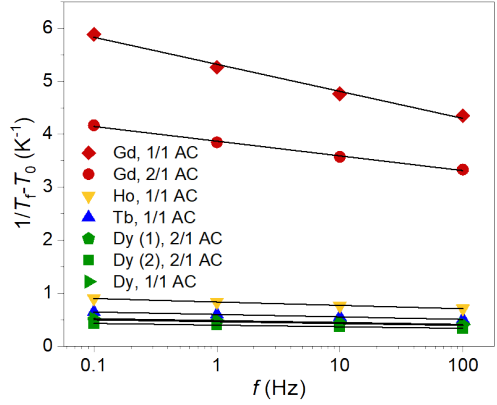

Fig. 12. variation of $1/(T_f - T_0)$ in the SG samples including 1/1 and 2/1 ACs as a function of measuring frequency $f$.

$2.19(0) - 2.64(5)$ in the Dy-contained ACs, and $1.26(3)$ in the Ho-contained 2/1 AC. These values satisfy the $T_f - T_0 \ll T_f$ condition in the Gd-contained ACs where the strong RKKY-type interaction is expected. For the non-Heisenberg SGs, though, $T_0 \ll T_f$ is satisfied implying a weak dominance of RKKY interaction mediated by itinerant electrons, which is also the case in Al-Pd-Mn iQC [64]. Note that the same approach has already been applied on some iQCs and ACs with SG behavior [33,43,64,66].

## IV. CONCLUSION

In the present article, we investigated magnetic properties of the Ga-Pd-RE (RE = Gd, Tb, Dy and Ho) 1/1 and 2/1 ACs in detail. The magnetic susceptibilities (both dc and ac) evidenced SG-like freezing behavior for most of the studied samples except for Ga-Pd-Tb 2/1 AC and Ga-Pd-Ho 1/1 AC. In the former, a long-range AFM order is noticed at 5.78 K. The empirical frustration parameter $|\theta_w/T_f|$ indicated lower frustration levels in the 2/1 ACs compared with their corresponding 1/1 ACs. This phenomenon was partly correlated to the malformed octahedron arrangement of RE sites and disorder-free environment in their nearest neighbour in the 2/1 ACs. The ac magnetic susceptibility results showed the least sensitivity of the Gd-contained SGs (Heisenberg systems) to the frequency change, while non-Heisenberg ones were very much responsive. This phenomenon was associated with the relatively weak energy of the spin-orbital coupling via CEF in $Gd^{3+}$ making it easier to be overcome during the spin rotation. Furthermore, the frequency-dependence of $T_f$ in SG samples was explained by means of the Vogel–Fulcher law.

As for the future works, one may consider examining the magnetic behavior of the Ga-Pd-Ho 2/1 AC below 1.8 K where establishment of long-range order seems to be likely. Another follow-up research could focus on comparative study of aging, relaxation and rejuvenation in Heisenberg and non-Heisenberg SGs in more detail.

## ACKNOWLEDGMENT

This work was supported by Japan Society for the Promotion of Science through Grants-in-Aid for Scientific Research (Grants No. JP19H05817, No. JP19H05818, No. JP19H05819, and No. JP21H01044) and JST, CREST Grant No. JPMJCR22O3, Japan.

…